\documentclass{article}
\usepackage{INTERSPEECH2020}

\usepackage{color,xcolor}
\usepackage{epsfig}
\usepackage{graphicx}

\usepackage{array}
\usepackage{booktabs}
\usepackage{colortbl}
\usepackage{multirow}
\usepackage{footnote}
\makesavenoteenv{tabular}
\makesavenoteenv{table}

\usepackage{amsmath,amsfonts,amsthm,amssymb,bm,upgreek,mathtools}
\usepackage[super]{nth}

\usepackage{changepage}
\usepackage{extramarks}
\usepackage{lastpage}
\usepackage{setspace}
\usepackage{soul}
\usepackage{xspace}

\usepackage{url}
\usepackage{hyperref}
\hypersetup{colorlinks=True, urlcolor=black}
\usepackage[numbers,sort]{natbib}

\usepackage{algorithm, algorithmic}
\usepackage{enumitem}
\usepackage{verbatim}
\usepackage{pifont}
\usepackage[acronyms]{glossaries}
\glsdisablehyper

\usepackage{lipsum}

\newcommand{\sect}[1]{Section~\ref{sec:#1}}
\newcommand{\sectdot}[1]{Sec.~\ref{sec:#1}}
\newcommand{\ssect}[1]{Section~\ref{ssec:#1}}
\newcommand{\ssectdot}[1]{Sec.~\ref{ssec:#1}}

\newcommand{\eqndot}[1]{Eqn.~(\ref{eqn:#1})}

\newcommand{\tbl}[1]{Table~\ref{tab:#1}}

\newcommand{\twotbl}[2]{Tables~\ref{tab:#1} and \ref{tab:#2}}

\newcommand{\ignore}[1]{}
\newcommand{\norm}[1]{\lVert#1\rVert}

\DeclareMathOperator*{\argmax}{arg\,max}

\makeatletter
\DeclareRobustCommand\onedot{\futurelet\@let@token\@onedot}
\def\@onedot{\ifx\@let@token.\else.\null\fi\xspace}

\def\eg{\emph{e.g}\onedot} 
\def\ie{\emph{i.e}\onedot}

\makeatother

\definecolor{MyDarkBlue}{rgb}{0,0.08,1}
\definecolor{MyDarkGreen}{rgb}{0.02,0.6,0.02}
\definecolor{MyDarkRed}{rgb}{0.8,0.02,0.02}
\definecolor{MyDarkOrange}{rgb}{0.40,0.2,0.02}
\definecolor{MyPurple}{RGB}{111,0,255}
\definecolor{MyRed}{rgb}{1.0,0.0,0.0}
\definecolor{MyGold}{rgb}{0.75,0.6,0.12}
\definecolor{MyDarkgray}{rgb}{0.66, 0.66, 0.66}

\def\presec{\vspace{0.0em}}
\def\postsec{\vspace{0.0em}}

\definecolor{spkA}{HTML}{1f77b4}
\definecolor{spkB}{HTML}{ff7f0e}
\definecolor{spkC}{HTML}{2ca02c}
\definecolor{spkD}{HTML}{d62728}
\definecolor{spkE}{HTML}{9467bd}

\newcommand{\argmaxof}[1]{\underset{#1}{\argmax}\:}
\newcommand\given[1][]{#1\vert}

\newcommand{\sscdvat}{\textnormal{\tiny{\sc cdvat}}}
\newcommand{\mathcd}{cd}
\newcommand{\labd}{\mathcal{D}_{l}}
\newcommand{\dpl}{N_l}
\newcommand{\unlabd}{\mathcal{D}_{ul}}
\newcommand{\dpul}{N_{ul}}
\newcommand{\loss}{\mathcal{L}}
\newcommand{\lossof}[3]{\loss\left(#1,#2,#3\right)}

\newcommand{\suploss}{l}
\newcommand{\suplossof}[2]{\suploss\!\left(#1,#2\right)}

\newcommand{\cdloss}{\mathcal{R}_{\sscdvat}}
\newcommand{\cdlossof}[3]{\cdloss\!\left(#1,#2,#3\right)}

\newcommand{\lcs}{\textnormal{LCS}}
\newcommand{\lcsof}[2]{\lcs\!\left(#1,#2\right)}
\newcommand{\pars}{\uptheta}
\newcommand{\parsfix}{\hat{\pars}}
\newcommand{\ex}{\mathbf{x}}
\newcommand{\pertbase}{r}
\newcommand{\vatpert}{\mathbf{\pertbase}_{\sscdvat}}
\newcommand{\pert}{\mathbf{\pertbase}}
\newcommand{\pgrad}{\nabla_{\pert}}
\newcommand{\plapl}{\nabla\nabla_{\pert}}
\newcommand{\pertatit}[1]{\mathbf{v}_{#1}}
\newcommand{\emb}{\mathbf{e}}
\newcommand{\embof}[2]{\emb\!\left(#1,#2\right)}
\newcommand{\embp}{\emb_{\pertbase}}
\newcommand{\cdof}[2]{\mathcd\!\left[#1,#2\right]}
\newcommand{\simpcd}[3]{\mathcd\!\left[#1,#2,#3\right]}
\newcommand{\eigof}[3]{\lambda_{1}\!\left(#1,#2,#3\right)}
\newcommand{\cdofeq}[2]{\frac{1}{2} - \frac{#1^T#2}{2 \norm{#1}\norm{#2}}}
\newcommand{\hess}{\mathbf{H}}
\newcommand{\hessof}[2]{\hess\!\left(#1,#2\right)}
\newcommand{\eigv}{\mathbf{u}}
\newcommand{\eigvof}[2]{\eigv\!\left(#1,#2\right)}
\newcommand{\unaryminus}{\scalebox{0.5}[1.0]{\( - \)}}
\DeclarePairedDelimiter\ceil{\lceil}{\rceil}

\newacronym{rnn}{RNN}{recurrent neural network}
\newacronym{ann}{ANN}{artificial neural network}
\newacronym{dnn}{DNN}{deep neural network}
\newacronym{mlp}{MLP}{Multi-layer perceptron}
\newacronym{asr}{ASR}{automatic speech recognition}
\newacronym{tdnn}{TDNN}{time-delay neural network}
\newacronym{fc}{FC}{fully-connected}
\newacronym{asoft}{ASoftmax}{angular softmax}
\newacronym{dnc}{DNC}{Discriminative Neural Clustering}
\newacronym{mha}{MHA}{multi-head attention}
\newacronym{vad}{VAD}{voice activity detection}
\newacronym{cpd}{CPD}{speaker change point detection}
\newacronym{ser}{SER}{speaker error rate}
\newacronym{der}{DER}{diarisation error rate}
\newacronym{uisrnn}{UIS-RNN}{unbounded interleaved-state \gls{rnn}}
\newacronym{std}{std}{standard deviation}
\newacronym{tsne}{t-SNE}{t-distributed stochastic neighbour embedding}
\newacronym{sc}{SC}{spectral clustering}
\newacronym{mdm}{MDM}{multiple distance microphone}
\newacronym{daug}{Diac-Aug}{\emph{Diaconis augmentation}}
\newacronym{cl}{CL}{curriculum learning}
\newacronym{pit}{PIT}{permutation invariant training}
\newacronym{gan}{GAN}{generative adversarial network}
\newacronym{vtln}{VTLN}{vocal tract length normalization}
\newacronym{sgd}{SGD}{stochastic gradient descent}

\newacronym{ssl}{SSL}{semi-supervised learning}
\newacronym{vat}{VAT}{virtual adversarial training}
\newacronym{cdvat}{CD-VAT}{cosine-distance virtual adversarial training}
\newacronym{wrt}{w.r.t}{with respect to}
\newacronym{mfcc}{MFCC}{mel-frequency cepstral coefficient}
\newacronym{eer}{EER}{equal error rate}
\newacronym{roc}{ROC}{receiver operating characteristic}
\newacronym{iss}{ISS}{inter-speaker separability}
\newacronym{isc}{ISC}{intra-speaker compactness}
\newacronym{plda}{PLDA}{probabilistic linear discriminant analysis}

\newacronym{gdpr}{GDPR}{General Data Protection Regulation}
\newacronym{ccpa}{CCPA}{California Consumer Privacy Act}

\title{Cosine-Distance Virtual Adversarial Training for Semi-Supervised Speaker-Discriminative Acoustic Embeddings}
\name{Florian L. Kreyssig \& Philip C. Woodland}
\address{Cambridge University Engineering Dept., Trumpington St., Cambridge, CB2 1PZ U.K.}
\email{\{flk24,pcw\}@eng.cam.ac.uk}

\begin{document}

\maketitle
\begin{abstract}
In this paper, we propose a \gls{ssl} technique for training \glspl{dnn} to generate speaker-discriminative acoustic embeddings (speaker embeddings). Obtaining large amounts of speaker recognition training data can be difficult for desired target domains, especially under privacy constraints. The proposed technique reduces requirements for labelled data by leveraging unlabelled data. The technique is a variant of \gls{vat}~\cite{miyatoVirtualAdversarialTraining2018} in the form of a loss that is defined as the robustness of the speaker embedding against input
perturbations, as measured by the cosine-distance. Thus, we term the technique \gls{cdvat}. In comparison to many existing \gls{ssl} techniques, the unlabelled data does not have to come from the same set of classes (here speakers) as the labelled data. The effectiveness of \gls{cdvat} is shown on the 2750+ hour VoxCeleb data set, where on a speaker verification task it achieves a reduction in \gls{eer} of 11.1\% relative to a purely supervised baseline. This is 32.5\% of the improvement that would be achieved from supervised training if the speaker labels for the unlabelled data were available.
\end{abstract}

\glsresetall
\presec
\section{Introduction}
\postsec
\label{sec:intro}

Speaker-discriminative acoustic embeddings (or just speaker embeddings) derived through deep learning techniques have become the state-of-the-art for learning speaker representations~\cite{snyder2017deep,snyder2018xvec} to be used for tasks such as speaker recognition, speaker verification or speaker diarisation~\cite{snyder2017deep,diez2019bayesian,li2019discriminative,Villalba2020sota}. Previously i-vectors~\cite{dehak2011ivec} based on factor analysis were widely used.

The neural networks used to generate speaker embeddings are typically trained on a speaker classification task, for which the input is the acoustic feature sequence of an utterance and the output is the speaker label of that utterance~\cite{variani2014spkver,snyder2017deep}. By taking the output of a layer of this neural network (often the penultimate layer) as an embedding, a fixed dimensional vector can be generated for any given input utterance. This vector is speaker-discriminative due to the training objective. It has been found that such speaker embeddings can
be used to discriminate between speakers that are not present in the training data.

The quality of these embeddings will improve with the amount of training data and with the number of speakers in the training data, assuming the data comes from the target domain. However, the acquisition of enough suitable speaker classification data for the exact conditions one desires can be difficult. This is especially true as the regulations around identifiable user data tighten\footnote{See \gls{gdpr} or \gls{ccpa}.}. Under these constraints it is useful to use audio data with associated speaker labels together with de-identified (unlabelled) data to train speaker embedding generators.

This paper proposes a method that enables \gls{ssl} of speaker embeddings. In comparison to many \gls{ssl} methods in machine learning, the proposed method does not assume that the labelled and unlabelled data comes from the same classes (here speakers). Therefore, a small amount of labelled data from a small number of speakers can be complemented by a larger amount of data from a large number of speakers.
The proposed method is a newly derived sibling to \gls{vat}~\cite{miyatoVirtualAdversarialTraining2018} which is an \gls{ssl} method for classification tasks. Vanilla \gls{vat} assumes the labelled and unlabelled data to come from the same set of classes. This paper, however, attempts to utilise unlabelled data that comes from a completely different set of classes (here speakers).

The proposed \gls{ssl} technique, termed \gls{cdvat}, works by adding an additional loss to the standard supervised training loss. The loss is defined as the cosine-distance between a speaker embedding generated for an utterance and the embedding generated for the same utterance, which was perturbed by an {\em adversarial} noise that maximally increases the cosine-distance to the original, unperturbed, embedding. The loss is computed for every data point in the labelled and unlabelled data sets, thus smoothing the embedding generator \gls{wrt} the input for all data points lying on the data manifold. It can, therefore, be seen as a regularisation technique that is informed by the unlabelled data, which constrains the neural network to learn embeddings that generalise well to unseen speakers.

This paper is organised as follows. \sect{cdvat} describes the \gls{cdvat} loss and how the adversarial noise is computed. In \sectdot{setup}, the experimental setup is described including the data sets and evaluation metrics used. In \sectdot{results} the experimental results are presented and \sectdot{conclusions} gives conclusions.
\glsreset{cdvat}
\presec
\section{Cosine-Distance VAT}
\postsec
\label{sec:cdvat}

In nature, the outputs of most systems are smooth \gls{wrt} spatial and temporal inputs~\cite{wahba1990spline}. Prior studies have confirmed that smoothing the output distribution of a classifier
(\ie, encouraging the classifier to output similar distributions)
against perturbations of the input can improve
its generalisation performance in \acrlong{ssl}~\cite{Sajjadi2016pert,Laine2016tempens,luo2017smooth,miyatoVirtualAdversarialTraining2018}. In the standard version of \gls{vat}~\cite{miyatoVirtualAdversarialTraining2018}~(the efficacy of which has been verified by~\cite{oliverRealisticEvaluation2018,zhai2019s4l}) an additive loss is introduced, which tries to smooth the categorical output distribution (measured by the KL-divergence) around every data point that lies on the data manifold. Here, \gls{vat} will be formulated on the level of the embedding layer rather than the output (classification) layer. The purpose of the proposed variant of \gls{vat} is to smooth the embedding generator in terms of the cosine-distance, termed \gls{cdvat}. \gls{cdvat} should be used together with an angular penalty loss (such as angular softmax~\cite{Liu2017sphereface}) in comparison to the standard cross-entropy loss. These types of losses are very popular for both
speaker verification~\cite{li2018angular,liu2019large,luu2020dropclass}
as well as face verification and identification~\cite{Liu2017sphereface,deng2019arcface}. When angular penalty losses are used during speaker classification training, the resulting embedding generator produces embeddings that are angularly discriminative \ie the cosine-distance between embeddings indicates if embeddings come from the same speakers.

\presec
\subsection{\texorpdfstring{\gls{cdvat}}{CD-VAT} loss}
\postsec

\Gls{cdvat} adds an additional loss $\cdloss$ (the \gls{cdvat} loss) to the supervised loss $\suplossof{\cdot}{\cdot}$ with the interpolation constant $\alpha$ and is computed on both the labelled data set $\labd$ (size $\dpl$) and the unlabelled data set $\unlabd$ (size $\dpul$). The combined loss $\loss$ is then used to train the parameters $\pars$ and in turn the embedding generator $\embof{\ex}{\pars}$.
\begin{equation}
    \lossof{\labd}{\unlabd}{\pars} = \suplossof{\labd}{\pars} + \alpha \cdlossof{\labd}{\unlabd}{\pars}\\
\end{equation}
The \gls{cdvat} loss, $\cdloss$, is the sum of local 
losses $\lcsof{\ex}{\pars}$ that are computed for each input feature sequence $\ex \in \labd, \unlabd$.
\begin{equation}
    \cdlossof{\labd}{\unlabd}{\pars} = \frac{1}{\dpl + \dpul} \sum_{\ex \in \labd, \unlabd} \lcsof{\ex}{\pars}
\end{equation}
%
The local cosine smoothness, $\lcsof{\ex}{\pars}$, is calculated in two steps.
First, a perturbation ($\vatpert$) to the input sequence $\ex$ is found. This perturbation is chosen to be an \textit{adversarial}\footnote{Note: no relationship to \glspl{gan}.} perturbation that maximally changes the embedding ($\embof{\ex\!+\! \vatpert}{\pars}$) of the input feature sequence, $\ex$, as measured by the cosine-distance ($\cdof{\cdot}{\cdot}$). $\epsilon$ is the maximum norm of $\vatpert$.
\begin{align}
    \vatpert =& \argmaxof{\pert; \norm{\pert}
    \leq \epsilon} \cdof{\embof{\ex}{\parsfix}}{\embof{\ex + \pert}{\parsfix}}\given[\Big]_{\parsfix=\pars}\label{eqn:advpert}\\
    \cdof{\mathbf{a}}{\mathbf{b}} =& \cdofeq{\mathbf{a}}{\mathbf{b}}
\end{align}
Second, $\lcsof{\ex}{\pars}$ is then the cosine-distance between the embedding of the (maximally) perturbed input sequence and the embedding for the unperturbed input sequence.
\begin{equation}
    \lcsof{\ex}{\pars} = \cdof{\embof{\ex}{\parsfix}}{\embof{\ex + \vatpert}{\pars}}\given[\Big]_{\parsfix=\pars}
\end{equation}
$\parsfix$ is the current setting for $\pars$ at a particular instant during optimisation \ie it is treated as a constant. The distinction between $\pars$ and $\parsfix$ is made because gradients of $\lcsof{\ex}{\pars}$ are only propagated back through the embedding generated with the input perturbation \ie $\embof{\ex + \vatpert}{\pars}$ and not $\embof{\ex}{\parsfix}$.
$\lcsof{\ex}{\pars}$ indicates how ``sensitive" the embedding of input $\ex$ is.

\presec
\subsection{Approximation of \texorpdfstring{$\vatpert$}{the virtual adversarial perturbation}}
\postsec

Given the adversarial perturbation $\vatpert$, the optimisation of the combined loss $\loss$ is straightforward, because the gradients of $\lcs$ \gls{wrt} $\pars$ are well defined\footnote{\eqndot{cdvatgrad} can be used.}. In this section a method for approximately finding $\vatpert$ is described. For simplicity, let: 
\begin{equation}
    \simpcd{\pert}{\ex}{\pars} = \cdof{\embof{\ex}{\parsfix}}{\embof{\ex + \pert}{\parsfix}}\given[\Big]_{\parsfix=\pars}
\end{equation}
$\simpcd{\pert}{\ex}{\pars}$ has a minimum of zero at $\pert\!=\!\mathbf{0}$. 
Therefore, the gradient \gls{wrt} $\pert$ is also zero at $\pert\!=\!\mathbf{0}$. Therefore, the second-order Taylor approximation of $\simpcd{\pert}{\ex}{\pars}$ is given by:
\begin{align}
    \simpcd{\pert}{\ex}{\pars} \approx \frac{1}{2} \pert^T \hessof{\ex}{\pars} \pert
\end{align}
where $\hessof{\ex}{\pars}\!=\! \plapl\simpcd{\pert}{\ex}{\pars} \given_{\pert=\mathbf{0}}$. For simplicity, let $\hess\!=\!\hessof{\ex}{\pars}$. Under this approximation $\vatpert$ emerges as the dominant eigenvector $\eigvof{\ex}{\pars}$ of $\hess$ with magnitude $\epsilon$ (see constraint from \eqndot{advpert}). This shows that \gls{cdvat} in effect penalises $\eigof{\pert}{\ex}{\pars}$, the largest eigenvalue of $\hess$:
\begin{align}
    \simpcd{\pert}{\ex}{\pars} \approx \frac{1}{2} \epsilon^{2} \eigof{\pert}{\ex}{\pars}
\end{align}
The dominant eigenvector, $\eigv$, of $\hess$ can be found using the standard power iteration method~\cite{kreyszig11} combined with finite differences. Let $\pertatit{0}$ be a randomly sampled vector that is not orthogonal to $\eigv$. Then the iterative calculation of
\begin{align}
    \pertatit{i+1} \leftarrow \overline{\hess\pertatit{i}}
\end{align}
causes $\pertatit{i}$ to converge to $\eigv$. The Hessian-vector product, $\hess\pertatit{i}$, can be approximated based on finite differences.
\begin{align}
    \pgrad \simpcd{\pert}{\ex}{\pars} \given_{\pert=\zeta \pertatit{i}} &\approx \pgrad \simpcd{\pert}{\ex}{\pars} \given_{\pert=\mathbf{0}} + \zeta \hess\pertatit{i}\\
    \hess\pertatit{i} 
%
    \approx \tfrac{1}{\zeta} \pgrad& \simpcd{\pert}{\ex}{\pars} \given_{\pert=\zeta \pertatit{i}}
\end{align}
Therefore, to obtain $\vatpert$ we can use the iterative procedure:
\begin{align}
    \vatpert &\approx \epsilon\cdot \pertatit{K}\\
    \pertatit{i+1} &= \frac{\mathbf{g}_{i+1}}{\norm{\mathbf{g}_{i+1}}}\\
    \mathbf{g}_{i+1} &= {\pgrad \simpcd{\pert}{\ex}{\pars} \given_{\pert=\zeta \pertatit{i}}}\label{eqn:calcnoise}
\end{align}
where $\pertatit{K}$ is the approximation of $\eigvof{\ex}{\pars}$ after $K$ power iterations and $\pertatit{0}$ is sampled uniformly on the unit-sphere. The value of $\zeta$ should be as small as possible to get the best estimate of $\hess\pertatit{i}$, but large enough not to cause numerical issues. Here, $\zeta$ is set to \mbox{0.005} in all our experiments\footnote{For our experiments this is a norm of \mbox{10e-6} per feature vector.}.

\mbox{$\mathbf{g}_{i+1}=\pgrad \simpcd{\pert}{\ex}{\pars}\given_{\pert=\zeta \pertatit{i}}$} is derived below:
\begin{align}
    \mathbf{g}_{i+1} &= \frac{\partial \simpcd{\pert}{\ex}{\pars}}{\partial \pert}\given[\Big]_{\pert=\zeta \pertatit{i}}\\
    &= \frac{\partial \embof{\ex+\pert}{\parsfix}}{\partial \pert}\frac{\partial \simpcd{\pert}{\ex}{\pars}}{\partial \embof{\ex+\pert}{\parsfix}}\given[\Bigg]_{\pert=\zeta \pertatit{i}}\label{eqn:bprop}\\
    \textnormal{let } \embof{\ex}{\parsfix} &= \emb \textnormal{ and } \embof{\ex+\pert}{\parsfix} = \embp\nonumber\\
    \frac{\partial \simpcd{\pert}{\ex}{\pars}}{\partial \embp} &= \frac{\unaryminus1}{2 \norm{\emb}}
    \cdot \frac{\partial}{\partial \embp}
    \left(\frac{\emb^T\embp}{\norm{\embp}}\right)\\
    =& \frac{\unaryminus1}{2 \norm{\emb}}
    \cdot
    \Big(
        \frac{1}{\norm{\embp}} \cdot \frac{\partial}{\partial \embp}\left(
            \emb^T\embp
        \right)\nonumber\\
    &+   \frac{\partial}{\partial \embp}\left(
            \frac{1}{\norm{\embp}}
        \right) \cdot \left(
            \emb^T\embp
        \right)
    \Big)\\
    =& \frac{\unaryminus1}{2 \norm{\emb}}
    \cdot
    \left(
        \frac{1}{\norm{\embp}} \cdot \emb
    -   \frac{{\embp}}{\norm{\embp}^{3}} \cdot \left(\emb^T\embp\right)\right)\label{eqn:cdvatgrad}
\end{align}
The pre-multiplication with $\frac{\partial \embof{\ex+\pert}{\parsfix}}{\partial \pert}$ in \eqndot{bprop} is equivalent to the standard back-propagation algorithm.

\noindent
To summarise, to obtain $\vatpert$ the required calculations are:
\begin{itemize}[leftmargin=0.2in]
    \item a forward pass to get $\embof{\ex}{\parsfix} = \emb$ for \eqndot{cdvatgrad}
    \item then for each power iteration:
    \begin{itemize}[leftmargin=0.1in]
        \item a forward pass to get $\embof{\ex + \pertatit{i}}{\parsfix} = \embp$ for \eqndot{cdvatgrad}
        \item a backward pass to get ${\pgrad \simpcd{\pert}{\ex}{\pars} \given_{\pert=\zeta \pertatit{i}}}$ for \eqndot{calcnoise}
    \end{itemize}
\end{itemize}

Our experiments on multiple data sets suggest that $K\!=\!1$ is sufficient, such that $\vatpert$ does not change significantly for further iterations as measured by the dot-product of consecutive $\pertatit{i}$. This single power iteration, however, increases $\simpcd{\pert}{\ex}{\pars}$ by up to a factor of \mbox{$10^4$} in comparison to just using $\pertatit{0}$ \ie the robustness of the embedding to the simple normalised Gaussian noise $\pertatit{0}$ is far larger than to the adversarial noise $\vatpert$.

\presec
\subsection{Related Work}
\postsec
\label{ssec:related}

Adversarial training was originally proposed by~\cite{szegedy2013intriguing}, where it was discovered that for image classification \glspl{dnn} are very vulnerable to input perturbations applied to an input in the direction to which the model's label assignment is the most sensitive, even when the perturbation is so small that human eyes cannot discern the perturbation. Such perturbed data points are also known as \textit{adversarial examples} and can be used as additional data points for training~\cite{goodfellow2014explaining}. For speaker verification~\cite{xuliadversarial} has investigated robustness to adversarial examples.

Commonly used \gls{ssl} techniques in speech processing include self-training. The classifier is first trained on the labelled data and then used to assign labels to the unlabelled data. The unlabelled data (possibly after filtering) with assigned labels is combined with the labelled data. The classifier is then trained on the enlarged set of labelled data.
The first successful applications of self-training were speech recognition~\cite{zavaliagkosUtilizingUntranscribedTraining1998,kempUNSUPERVISEDTRAININGSPEECH1998,lamelLightlySupervisedUnsupervised2002,chanImprovingBroadcastNews2004} and word-sense
disambiguation~\cite{yarowskyUnsupervisedWordSense1995}.
For \gls{ssl} of speaker embeddings self-training cannot be used if the unlabelled data does not come from the same set of speakers as the labelled data. 

Self-supervised learning for speaker embeddings was used in~\cite{stafylakis2019self} where speaker embeddings were trained via reconstructing the frames of a target speech segment, given the inferred embedding of another speech segment of the same utterance. In comparison to \gls{cdvat} their method needs the training data to be segmented into utterances that each belong to one speaker. A purely unsupervised approach for speaker embeddings that is also based on a reconstruction loss is proposed in~\cite{Peng2019MixtureFA}. Though these unsupervised approaches demonstrate impressive results for generating unsupervised speaker embeddings, they still require supervised data to train the \gls{plda} back-end used for speaker verification.

\gls{cdvat} notably improves the \gls{isc}. In~\cite{le2018robust} the \gls{isc} is directly optimised by adding a supervised loss to the triplet loss that is otherwise used.

Related to \gls{vat} are consistency-based \gls{ssl} methods, such as the mean teacher method~\cite{tarvainen2017mean} which was applied to audio command classification by~\cite{lu2019semi}. Another popular method to reduce labelling effort is to more effectively exploit the existing labelled corpus through data augmentation~\cite{snyder2018xvec,yamamoto2019speaker}.



\glsreset{eer}
\presec
\section{Experimental Setup}
\postsec
\label{sec:setup}

Experiments were designed to evaluate the effect of \gls{cdvat} on general speaker verification performance, while also more directly testing the effect of \gls{cdvat} on \acrlong{isc} and \acrlong{iss}.

\presec
\subsection{Data Sets}
\postsec
\label{ssec:data}

Two data sets, VoxCeleb1 (dev+test) and VoxCeleb2 (dev+test) are used to train the models and evaluate speaker verification performance. The VoxCeleb data sets consist of utterances that were obtained from Youtube videos and automatically labelled using a visual speaker recognition pipeline.
The VoxCeleb2 (dev+test) data set, together with the dev portion of the VoxCeleb1 data set, is used for training. For evaluation, the test portion of VoxCeleb1 is used. The combined train set consists of more than 2750 hours of data from 7323 
speakers. The evaluation set consists of 4874 utterances from 40 speakers, for which the official speaker verification list of 37720 utterance pairs is used. More information about the data is contained in \tbl{datasets}.


\begin{table}[!htb]
    \centering
    \begin{tabular}{l|r|r}
        \toprule
        Title &  train & test\\\midrule
        \# Speakers   & 7323 & 40 \\
        \# Videos & 172299 & 677\\
        \# Utterances  & 1276888 & 4874 \\
        Avg. Utterance Len. & 7.85 sec & 8.25 sec\\
        \bottomrule
    \end{tabular}
    \vspace*{0.7em}
    \caption{The train data is the combination of VoxCeleb2 (dev+test) and the development portion of VoxCeleb1. The test data is the test portion of VoxCeleb1 for which 37720 utterance pairs are the verification list.}
    \label{tab:datasets}
    \vspace*{-1.9em}
\end{table}
\noindent
The system input features are 30-d \glspl{mfcc}. The \glspl{mfcc} are extracted (using HTK~\cite{youngHTK}) using 25ms windows with 10ms frame increments from 30 filterbank channels. No \gls{vtln} was applied and $c_0$ is used instead of energy. These inputs were normalised
at the utterance level for mean and globally for
variance. No data augmentation was used for these experiments.

\presec
\subsection{Model Architecture}
\postsec
\label{ssec:model}

In our model, utterance-level speaker embeddings are created by averaging multiple $L_2$-normalised window-level embeddings. The input window to the window-level embedding generator is around 2 seconds (213 frames, [-106,+106]) long. The shift between windows is just under 100 frames (see details below).
The embedding generator uses a \gls{tdnn}~\cite{peddintiTimeDelayNeural,kreyssig2018TDNN} with a total input context of [-7,+7], which is shifted from \{-99\} to \{+99\} with shifts of 6 frames (resulting in the overall input window of [-106,+106]). The 
34 output vectors of the \gls{tdnn} are combined using the self-attentive layer proposed in~\cite{sun2019diarselfatt}. This is followed by a linear projection down to the embedding size, 
which is then the window-level embedding. The TDNN structure resembles the one used in the x-vector models \cite{snyder2018xvec} (\ie \gls{tdnn}-layers with the following input contexts: [-2,+2], followed by \{-2,0,+2\}, followed by \{-3,0,+3\}, followed by \{0\}). The first three \gls{tdnn}-layers have a size of 512, the third a size of 256 and the embedding size is 32.

An utterance of length $T$ fits $N=\ceil{\frac{T-213}{100}}$ full windows (at shifts of 100 frames) plus another window if padding were used (\eg replication padding to $213\!+\!N*100$ frames). To avoid padding, shifts of $\frac{T-213}{N}$ are used (\ie slightly under 100 frames). The resulting indices are rounded to the nearest integer. For utterances shorter than 213 frames, the window is aligned to the centre of the utterance and replication padding used.

\presec
\subsection{Training}
\postsec
\label{ssec:Training}

\gls{cdvat} was implemented in HTK~\cite{youngHTK} with which all models were trained in conjunction with PyHTK~\cite{zhang2019pyhtk}. For training, the window-level embedding is classified into the different speakers. The training objective for supervised training is angular softmax~\cite{Liu2017sphereface} with $m\!=\!1$. The embedding generator is optimised using \gls{sgd} with momentum, and weightdecay was used for regularisation. The learning rate scheduler is NewBob. The batch size used for the supervised loss, $\suplossof{\cdot}{\cdot}$, was 200 except for the model trained on the entire data set for which a batch size of 400 was used. The batch size used for the \gls{cdvat} loss, $\cdloss$, was 800 \ie four times higher. The interpolation coefficient $\alpha$ was set to 0.4 and the norm of the adversarial perturbation $\epsilon$ was set to 13. The model was trained directly on the combined loss, $\loss\!\left(\cdot\right)$, \ie not pre-trained on the purely supervised loss.

For the experiments two partitions into labelled and unlabelled were created. For one 220k utterances are labelled and 440k utterances for the other. The remaining 1057k and 837k utterances, respectively, form the unlabelled dataset. The 220k and 440k utterances are chosen from the top of the utterance list sorted by official utterance name. Of the labelled dataset 20k and 40k utterances, respectively, form the validation set.

\presec
\subsection{Evaluation Criteria (\texorpdfstring{\acrshort{eer}}{EER}, \texorpdfstring{\acrshort{isc}}{ISC}, \texorpdfstring{\acrshort{iss})}{ISS}}
\postsec
\label{ssec:evaluation}\glsreset{eer}\glsreset{isc}\glsreset{iss}
The main evaluation criterion is the speaker verification \gls{eer}. First, utterance embeddings are formed for each utterance by averaging the window embeddings, where the windows are based on the shifts described at the end of \ssectdot{model}. The scores necessary for the \gls{roc} curve used in the \gls{eer} calculation are the cosine-distances between the embeddings of utterance pairs, in comparison to the otherwise commonly used \gls{plda} backend\footnote{\cite{luu2020dropclass,li2018angular} are other publications that experiment with direct cosine-scoring for speaker verification.}. The \gls{roc} curve is built using scikit-learn~\cite{scikit2011learn}.

Speaker-discriminative acoustic embeddings should have two qualities, \gls{isc} \ie how close to each other are the embeddings of a single speaker and \gls{iss} \ie how far apart are the embeddings of different speakers. To give further insight about the embeddings generated from our models we attempt to give a measure of these two qualities. First all utterance embeddings belonging to the same speaker are collected and the centroid calculated (one per speaker). For the \gls{iss} the average pairwise distance between the centroids of all speakers is found. For the \gls{isc}, all utterance embeddings belonging to the same speaker are collected and the average cosine-distance to the centroid of that speaker found and the average of those per-speaker scores calculated.
\presec
\section{Experimental Results}
\postsec
\label{sec:results}

\twotbl{EER}{ISC} show the results of applying \gls{cdvat} on the VoxCeleb data set. The supervised baseline model trained on 200k utterances of labelled data achieves an \gls{eer} of 8.32\%. The use of \gls{cdvat} reduces the \gls{eer} by 11.1\% relative down to an \gls{eer} of 7.40\%. This represents an \gls{eer} recovery of 32.5\% \ie we achieve 32.5\% of the reduction in \gls{eer} that we would get from pure supervised training if we had the speaker labels for the unlabelled part of the training data (such a model has an \gls{eer} of 5.52\%). This error rate recovery is similar to those seen in other areas of machine learning. For instance, \cite{manoharSemisupervisedMaximumMutual2015} presents a word error rate recovery of 37\% for \gls{ssl} (minimum entropy training) of a DNN-HMM speech recogniser without additional language modelling data. At the same time \gls{ssl} for speaker embeddings presents additional challenges as for larger numbers of speakers class overlap can exist. The information content of unlabelled examples decreases as classes overlap as shown by~\cite{castelliRelativeValueLabeled1996,oneillNormalDiscriminationUnclassified1978}.

\begin{table}[t!]
    \centering
    \begin{tabular}{c|cc|c|r}
        \toprule
        System & Utts $\labd$ & \#Speakers & Utts $\unlabd$ & EER\\\midrule
        Sup 1 & \multirow{2}{*}{200k}  & \multirow{2}{*}{1249} & - & 8.32\%\\
        CDVAT 1 &   &  & 1057k & 7.40\%\\\midrule
        Sup 2 & \multirow{2}{*}{400k}  & \multirow{2}{*}{2504} & - & 6.85\%\\
        CDVAT 2 &  &  & 837k & 6.46\%\\\midrule
        Sup 3 & 1277k  & 7323 & - & 5.52\%\\
        \bottomrule
    \end{tabular}
    \vspace*{0.8em}
    \caption{Evaluation of \acrshort{cdvat} on the VoxCeleb dataset. The evaluation criterion \acrshort{eer} is explained in \ssect{evaluation}. For \acrshort{eer} a lower value is better. $\labd$ is the labelled data set and $\unlabd$ is the unlabelled data set.}
    \label{tab:EER}
    \vspace*{-2.5em}
\end{table}

The supervised baseline model trained on 400k utterances of labelled data achieves an \gls{eer} of 6.85\%. The use of \gls{cdvat} reduces the \gls{eer} by 5.7\% relative down to an \gls{eer} of 6.46\%. This represents an \gls{eer} recovery of 29.3\%.

\begin{table}[htb!]
    \centering
    \begin{tabular}{c|cc|c|r|r}
        \toprule
        System & Utts $\labd$ &  Utts $\unlabd$ &ISC & ISS\\\midrule
        Sup 1 & \multirow{2}{*}{220k}  & - & 0.13 & 0.38\\
        CDVAT 1 &  & 1057k & 0.09 & 0.36\\\midrule
        Sup 2 & \multirow{2}{*}{440k} & -& 0.14 & 0.40\\
        CDVAT 2 &  & 837k & 0.09 & 0.38\\\midrule
        Sup 3 & 1277k & - & 0.14 & 0.46\\
        \bottomrule
    \end{tabular}
    \vspace*{0.8em}
    \caption{Evaluation of \acrshort{cdvat} on the VoxCeleb dataset. The evluation criteria \acrshort{isc} and \acrshort{iss} are explained in \ssect{evaluation}. For \acrshort{isc} a lower value is better. For \acrshort{iss} a higher value is better.}
    \label{tab:ISC}
    \vspace*{-1.5em}
\end{table}

Furthermore, the \gls{isc}, which is very closely related to the \gls{cdvat} smoothing loss is reduced for the 200k and the 400k models by 31\% and 36\% respectively. However, at the same time the \gls{iss} is also slightly reduced. This shows one disadvantage of \gls{cdvat}, which is that it also brings the embeddings of all utterances closer together. To put these values into perspective, the threshold of the cosine-scoring used to obtain the \gls{eer} is between 0.42 and 0.48 for the systems trained.
\glsreset{cdvat}
\glsreset{isc}

\presec
\section{Conclusions}
\postsec
\label{sec:conclusions}

We have presented \gls{cdvat}, a method that allows for semi-supervised training of speaker-discriminative acoustic embeddings without the requirement that the set of speakers is the same for the labelled and the unlabelled data. It is shown that \gls{cdvat} can improve speaker verification performance on the VoxCeleb data set over a purely supervised baseline. The proposed method recovers 32.5\% of the \gls{eer} improvement that is obtained when speaker labels are available for the unlabelled data. \gls{cdvat} also significantly improves the \gls{isc} of the speaker embeddings. At the same time, however, the computational cost of \gls{cdvat} is twice as high (per data point) as supervised training and two new hyper-parameters, that need to be tuned, are introduced.

\presec
\section{Acknowledgements}
\postsec
Florian Kreyssig is funded by an EPSRC Doctoral Training Partnership Award.

\vfill
\newpage
\pagebreak

\presec
\section{References}
\vspace{-0.5em}
\begingroup
\renewcommand{\section}[2]{}
\bibliographystyle{CAMbib}
{\eightpt\bibliography{refs}
}
\endgroup

\end{document}